\documentclass[
reprint,
superscriptaddress,
amsmath,amssymb,
aps,
prl,
longbibliography,
lengthcheck,
showpacs
]{revtex4-1}

\usepackage{graphicx}
\usepackage{dcolumn}
\usepackage{bm}
\usepackage{hyperref}

\begin{document}


\title{Mass measurements of very neutron-deficient Mo and Tc isotopes and their impact on $rp$ process  nucleosynthesis}

\author{E.~Haettner}
\email[]{emma.k.haettner@physik.uni-giessen.de}
\affiliation{II.~Physikalisches Institut, Justus-Liebig-Universit\"at, 35392 Gie\ss{}en, Germany}
\affiliation{GSI Helmholtzzentrum f\"ur Schwerionenforschung, 64291 Darmstadt, Germany}

\author{D.~Ackermann}
\affiliation{GSI Helmholtzzentrum f\"ur Schwerionenforschung, 64291 Darmstadt, Germany}

\author{G.~Audi}
\affiliation{CSNSM-IN2P3-CNRS, 91405 Orsay-Campus, France}

\author{K.~Blaum}
\affiliation{Max-Planck-Institut f\"ur Kernphysik, 69117 Heidelberg, Germany}

\author{M.~Block}
\affiliation{GSI Helmholtzzentrum f\"ur Schwerionenforschung, 64291 Darmstadt, Germany}

\author{S.~Eliseev}
\altaffiliation[Present address: ]{Max-Planck-Institut f\"ur Kernphysik, 69117 Heidelberg, Germany}
\affiliation{GSI Helmholtzzentrum f\"ur Schwerionenforschung, 64291 Darmstadt, Germany}

\author{T.~Fleckenstein}
\affiliation{II.~Physikalisches Institut, Justus-Liebig-Universit\"at, 35392 Gie\ss{}en, Germany}

\author{F.~Herfurth}
\affiliation{GSI Helmholtzzentrum f\"ur Schwerionenforschung, 64291 Darmstadt, Germany}

\author{F.~P.~He\ss{}berger}
\affiliation{GSI Helmholtzzentrum f\"ur Schwerionenforschung, 64291 Darmstadt, Germany}

\author{S.~Hofmann}
\affiliation{GSI Helmholtzzentrum f\"ur Schwerionenforschung, 64291 Darmstadt, Germany}

\author{J.~Ketelaer}
\altaffiliation[Present address: ]{Max-Planck-Institut f\"ur Kernphysik, 69117 Heidelberg, Germany}
\affiliation{Institut f\"ur Physik, Johannes Gutenberg-Universit\"at, 55128 Mainz, Germany}

\author{J.~Ketter}
\affiliation{Max-Planck-Institut f\"ur Kernphysik, 69117 Heidelberg, Germany}

\author{H.-J.~Kluge}
\affiliation{GSI Helmholtzzentrum f\"ur Schwerionenforschung, 64291 Darmstadt, Germany}

\author{G.~Marx}
\affiliation{Institut f\"ur Physik, Ernst-Moritz-Arndt-Universit\"at, 17487 Greifswald, Germany}

\author{M.~Mazzocco}
\affiliation{Dipartimento di Fisica and INFN Sezione di Padova, 35131 Padova, Italy}

\author{Yu.~N.~Novikov}
\affiliation{GSI Helmholtzzentrum f\"ur Schwerionenforschung, 64291 Darmstadt, Germany}
\affiliation{Petersburg Nuclear Physics Institute, 188300 Gatchina, St.~Petersburg, Russia}

\author{W.~R.~Pla\ss{}}
\affiliation{II.~Physikalisches Institut, Justus-Liebig-Universit\"at, 35392 Gie\ss{}en, Germany}
\affiliation{GSI Helmholtzzentrum f\"ur Schwerionenforschung, 64291 Darmstadt, Germany}

\author{S.~Rahaman}
\altaffiliation[Present address: ]{LANL, Physics Division, P-23, MS-H803, Los Alamos 87545, USA}
\affiliation{Department of Physics, University of Jyv{\"a}skyl{\"a}, 40014 Jyv{\"a}skyl{\"a}, Finland}

\author{T.~Rauscher}
\affiliation{Department of Physics, University of Basel, 4056 Basel, Switzerland}

\author{D.~Rodr\'{i}guez}
\affiliation{Universidad de Granada, 18071 Granada, Spain}

\author{H.~Schatz}
\affiliation{Department of Physics and Astronomy, NSCL, and JINA, Michigan State University, East Lansing, USA}

\author{C.~Scheidenberger}
\affiliation{II.~Physikalisches Institut, Justus-Liebig-Universit\"at, 35392 Gie\ss{}en, Germany}
\affiliation{GSI Helmholtzzentrum f\"ur Schwerionenforschung, 64291 Darmstadt, Germany}

\author{L.~Schweikhard}
\affiliation{Institut f\"ur Physik, Ernst-Moritz-Arndt-Universit\"at, 17487 Greifswald, Germany}

\author{B.~Sun}
\affiliation{II.~Physikalisches Institut, Justus-Liebig-Universit\"at, 35392 Gie\ss{}en, Germany}
\affiliation{School of Physics and Nuclear Energy Engineering, Beihang University, 100191 Beijing, PR China}

\author{P.~G.~Thirolf}
\affiliation{Fakult\"at f\"ur Physik, Ludwig-Maximilians-Universit\"at, 85748 Garching, Germany}

\author{G.~Vorobjev}
\affiliation{GSI Helmholtzzentrum f\"ur Schwerionenforschung, 64291 Darmstadt, Germany}
\affiliation{Petersburg Nuclear Physics Institute, 188300 Gatchina, St.~Petersburg, Russia}

\author{M.~Wang}
\affiliation{Institute of Modern Physics, Lanzhou, PR China}

\author{C.~Weber}
\altaffiliation[Present address: ]{Fakult\"at f\"ur Physik, Ludwig-Maximilians-Universit\"at, 85748 Garching, Germany}
\affiliation{Department of Physics, University of Jyv{\"a}skyl{\"a}, 40014 Jyv{\"a}skyl{\"a}, Finland}


\date{\today}

\begin{abstract}
The masses of ten proton-rich nuclides, including the $N$=$Z$+1 nuclides $^{85}$Mo and $^{87}$Tc, were measured with the Penning trap mass spectrometer SHIPTRAP. Compared to the Atomic Mass Evaluation 2003 a systematic shift of the mass surface by up to 1.6 MeV is observed causing significant abundance changes of the ashes of astrophysical X-ray bursts. Surprisingly low $\alpha$ separation energies for neutron-deficient Mo and Tc are found, making the formation of a ZrNb cycle in the $rp$ process possible. Such a cycle would impose an upper temperature limit for the synthesis of elements beyond Nb in the $rp$ process. 
\end{abstract}


\maketitle

The astrophysical rapid proton capture process ($rp$ process) \cite{Wal81} is a reaction sequence involving very neutron-deficient nuclides near the proton drip line, possibly up to mass number $\simeq$100. It is thought to cause astronomically observed X-ray bursts by repeated thermonuclear explosions in a thin, proton-rich fuel layer on the surface of a mass accreting neutron star. The nuclear energy generation in a burst and the composition of the burst ashes need to be reliably predicted to understand the properties of bursts and their impact on the structure of the neutron star crust, which is related to neutron star cooling \cite{Gup07}. The $rp$ process may also produce the neutron-deficient $^{92,94}$Mo and $^{96,98}$Ru isotopes, whose origin cannot be understood in the framework of standard nucleosynthesis \cite{Arn03}. Although it is not obvious that in X-ray bursts sufficient amounts of material escape the neutron star surface to contribute to galactic nucleosynthesis and though large $rp$ process contributions may not be allowed by observed abundance patterns in meteorites \cite{Dau03}, such ejection is possible in principle \cite{Wei06}. 
Rapid proton captures on proton-rich nuclei are also an essential part of the $\nu p$ process occurring in the deepest, still ejected layers of core-collapse supernovae \cite{Fro06,Pru06}, where interactions with neutrinos come into play as well. The process, however, depends sensitively on the astrophysical conditions and on the explosion mechanism. Therefore the reaction path is not uniquely defined, though calculations indicate that it can reach neutron-deficient isotopes in the Mo, Ru region and beyond. The contribution of these processes to the origin of neutron-deficient isotopes in nature remains an open question. Improving the nuclear physics input is an important step to reliably calculate the isotopic production for a given astrophysical model. 

Nuclear masses strongly affect rapid proton capture processes \cite{Cle03,Sch06a} as the low proton separation energies ($S_\mathrm{p}$) encountered near the proton drip line determine where photodisintegration halts the sequence of proton captures and requires a $\beta^+$ decay, or a (n,p) reaction in case of the $\nu p$ process, for the process to proceed. While much progress has been made in recent years with precision mass measurements of proton-rich nuclides using Penning traps, see e.g. \cite{Kankainen_2006, Weber_2008,Kan08,Fal08,Breitenfeldt09} and references therein, the masses of crucial nuclides for model calculations beyond $A$$\approx$78 have not been measured yet. Thus, one has to rely on the extrapolations of the Atomic Mass Evaluation 2003 (AME03) \cite{AME2003} and on mass models. In this letter direct high-precision mass measurements of nuclides on the $rp$ process path are presented. 

At the velocity filter SHIP \cite{Hofmann_2000} super-heavy elements as well as nuclides close to the proton drip line are produced in fusion-evaporation reactions. 
For mass measurements, the exotic nuclides are transferred to the Penning trap mass spectrometer SHIPTRAP \cite{Block_2007}. As will be demonstrated in this work, the combination of SHIPTRAP with  the high beam intensities and superb separation capabilities of SHIP provides a unique reach for highly accurate mass measurements of extremely proton-rich nuclides. 
The products delivered by SHIP are stopped in a helium-filled gas cell and extracted into an RF quadrupole. There, the ions are cooled and transferred in bunches to the Penning trap system. In the first Penning trap isobaric separation is carried out by mass-selective buffer-gas cooling, and in the second trap accurate mass measurements are performed using the time-of-flight ion-cyclotron-resonance technique \cite{Graeff_1980}.

In the present experiment fusion-evaporation products in the element range from Rb to Tc were produced with a $^{36}$Ar primary beam of $4\times10^{12}$ particles/second at energies of 5.0 and 5.9 MeV/u reacting with a $^{54}$Fe target with an areal weight of 0.45 mg/cm$^{2}$, which has been enriched to 99\%. 
The helium pressure in the gas cell was varied between 50 and 60 mbar. The mass measurements were performed using excitation times between 0.1 and 2.4 s. Mostly an excitation time of 1.2 s was used,
which corresponds to a mass resolving power (FWHM) in excess of 10$^{6}$. The masses of ten nuclides $^{81}$Rb$^{m}$, $^{80,81,84}$Sr, $^{86}$Zr, $^{85}$Nb, $^{85,86,87}$Mo and $^{87}$Tc were measured with accuracies ranging from $4.6\times10^{-8}$ to $2\times10^{-7}$ using $^{85}$Rb as reference mass. Details of the experiment and the data analysis procedure will be the subject of a forthcoming publication.

The results of the mass measurements are given in table \ref{tab:Measurements}. The masses of $^{81}$Rb$^{m}$ and $^{80,81,84}$Sr were measured and found to be in agreement with the AME03. The more exotic nuclides, however, show a significant shift of the mass surface towards less bound nuclides as compared to the AME03 and thus follow the trend observed for Nb, Tc, Ru and Rh nuclides in previous experiments at SHIPTRAP and JYFLTRAP \cite{Kankainen_2006,Weber_2008}. 
The masses of $^{85}$Mo and $^{87}$Tc were measured for the first time and mass excesses are found to 
differ from the AME03 extrapolations by 1600 keV and  1400 keV, respectively, well outside of the 
300 keV extrapolation uncertainty given in AME03.
Moreover,  for $^{86,87}$Mo, which had only been indirectly measured before \cite{AME2003}, 
the shift in the mass excess values exceeds the previously given experimental uncertainties by several standard deviations.
The masses of $^{86}$Zr and $^{85}$Nb also deviate from previous measurements listed in AME03 but agree with the values obtained at JYFLTRAP \cite{Kankainen_2006}. 

\begin{table*}
\caption{\label{tab:Measurements}Measured frequency ratios (col. 2) of the nuclides investigated (col. 1) with the corresponding mass excess values based on a $^{85}$Rb mass of 84.911789738(12) u \cite{AME2003} (col. 3), as well as the corresponding AME03 values (col. 4), differences (col. 5), AMEup, (col. 6), $S_\mathrm{p}$ from AMEup (col. 7) and AME03 (col. 8) and their differences (col. 9). The symbol "\#" indicates an extrapolated value. }
\begin{ruledtabular}
\begin{tabular}{l|l|cccc|ccc}
 &  & \multicolumn{4}{c|}{Mass excess (keV)} & \multicolumn{3}{c}{Proton separation energy (keV)}\\
Nuclide & Frequency ratio $r$  &this work& AME03 & Difference & AMEup & AMEup & AME03 & Difference \\ \hline
$^{81}$Rb$^{m}$&0.95297789(11)&-75373.1(8.6)\footnote{The ground and isomeric states were not resolved. It is assumed that the isomeric state was populated, since it is favored by the reaction mechanism.}&-75369(6)&-4(10)&-75370.2(4.9)&&&\\
$^{80}$Sr&0.941264853(50)&-70313.4(3.9)&-70308(7)  &-5(8) &-70311.5(3.5)&6797.5(6.9)&6794(9)&4(11)\\
$^{81}$Sr&0.953026492(44)&-71528.6(3.5)&-71528(6)  &-1(7) &-71528.1(3.1)&6641.6(3.6)&6644(9)&-2(10)\\
$^{84}$Sr&0.988242201(51)&-80648.4(4.0)&-80644(3)  &-4(5) &-80649.1(1.3)&8863.1(6.1)&8858(7) &5(9)\\
$^{86}$Zr&1.011830049(54)&-77971.7(4.3)&-77800(30) & -170(30) &-77968.8(3.6)&7416(19)&7250(40)&170(40)\\
$^{85}$Nb&1.00200869(52)&-66279.7(4.1)&-67150(220)& 870(220) &-66279.7(4.1)&2147.7(7.0)&2950\#(300\#)&-800(300)\\
$^{85}$Mo&1.00031175(20) &-57510(16)&-59100\#(280\#)&1590(280)&-57510(16)&3778\#(300\#)&4510\#(410\#)&-730(510)\\
$^{86}$Mo&1.012005301(48)&-64110.3(3.8)&-64560(440)&450(440)  &-64110.2(3.7)&5119.5(5.5)&4700(490)&420(490)\\
$^{87}$Mo&1.023747251(49)&-66882.8(3.9)&-67690(220)&810(220)  &-66882.9(3.9)&5038.6(6.9) &5160(240)&-120(240)\\
$^{87}$Tc&1.023863477(53)&-57690.0(4.2)&-59120\#(300\#)&1430(300)&-57690.0(4.2)&868.8(5.6)&1860\#(530\#)&-990(530)\\
\end{tabular}
\end{ruledtabular}
\end{table*}


In order to investigate the influence of the new mass values on astrophysical models, a consistent mass surface is required without artificial steps in the separation energies as would occur if only individual mass values were replaced. For this reason, the new experimental mass data were added to the Atomic Mass Evaluation 
together with further recent experimental work \cite{Kankainen_2006,Weber_2008,Kan08,Fal08}. Taking into account 
the entire set of experimental masses in this region, adjusted mass values were obtained following the 
procedure employed in \cite{Wapstra_2003}. For the nuclides measured in this work these values are given in table \ref{tab:Measurements} column 6. 
Additionally, a consistent mass surface requires updated mass extrapolations due to the large mass shifts found. Thus, a new local extrapolation in the region $A$=80-95 was performed using the method and programs of \cite{Wapstra_2003}. Together with the new set of evaluated experimental data and the previously reported AME03 extrapolated mass values for $A$$<$80 and $A$$>$95 these data form an updated mass data set, which is labeled "AMEup" in this Letter.

\begin{figure}
\includegraphics[angle=0,width=80mm]{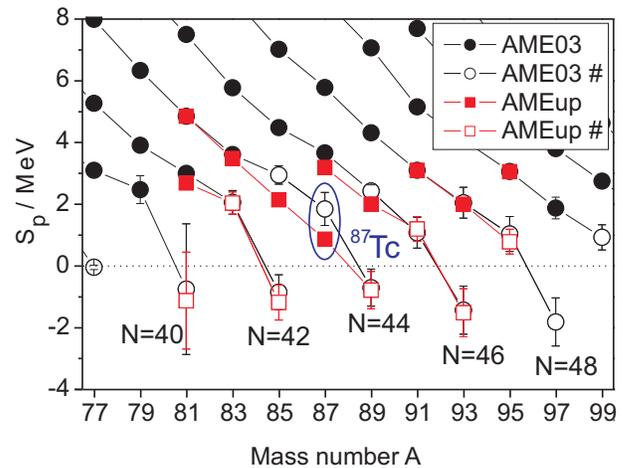}
\caption{\label{fig:S1p} (Color online) Proton separation energies ($S_\mathrm{p}$) of nuclides with odd $Z$ and even $N$ according to AME03 and AMEup. Filled symbols indicate measured values, open symbols indicate $S_\mathrm{p}$ where at least one of the two masses is extrapolated (\#).}
\end{figure}

Even though the masses generally are shifted in the same direction, the AMEup table does also show changes in the separation energies compared to AME03. 
In particular, the $S_\mathrm{p}$ for $^{87}$Tc determined in this work experimentally for the first time is 868(6) keV, about 1000 keV lower than the previous extrapolated value of 1860(530) keV. 
Generally, one observes lower $S_\mathrm{p}$ for the updated mass surface and a more linear trend (Fig.~\ref{fig:S1p}). 

To explore the impact of the new masses on the $rp$ process in X-ray bursts, reaction network calculations using an X-ray burst model \cite{Schatz_2001} were carried out. 
This single-zone model reproduces nucleosynthesis and energy generation reasonably well when compared to multi-zone burst models but its computational speed allows one to explore nuclear physics dependencies in detail.
The baseline calculation uses the nuclear masses of the AME03 and calculated Coulomb mass shifts \cite{Brown_2002} for nuclides beyond $N$=$Z$. The results are compared to network calculations based on the AMEup, combined with the same Coulomb mass shifts.

\begin{figure}
\includegraphics[angle=0,width=80mm]{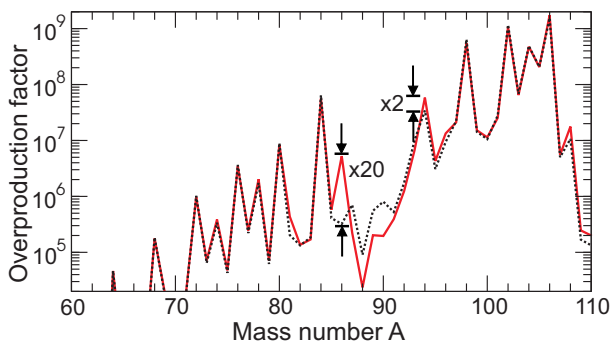}
\caption{\label{fig:abundance}(Color online) Calculated overproduction factors (produced abundance divided by solar system abundance) after an X-ray burst for the AME03 (dotted line) and the AMEup (solid line) mass sets. It has been assumed that all nuclei have decayed into the first stable isotope in each mass chain. 
The ratio of the overproduction factor for a given isotope to the highest factor found for any isotope (10$^9$) provides an upper limit for the contribution of X-ray bursts to the solar system abundance of that isotope. }


\end{figure}

The resulting final abundances show large differences between AME03 and AMEup in the region of $A$=86-96. The largest change is found for $A$=86 where the abundance increases by a factor of 20 (Fig.~\ref{fig:abundance}) due to the unexpected decrease in $S_\mathrm{p}$ of $^{87}$Tc. The lower $S_\mathrm{p}$ of $^{87}$Tc changes the ($\gamma$,p) and (p,$\gamma$) rates such that the $\beta$ decay branch for nuclides with mass number A=86 is strengthened. This change by a factor of 20 is by far the largest observed for abundances produced in $rp$ process network calculations since the AME 2003 evaluation. 
It demonstrates that nuclear physics  uncertainties can be larger than estimated and can introduce 
large uncertainties in nucleosynthesis model calculations.
Also the final abundance of $A$=94 increases by a factor of two. For mass numbers between 86 and 94 the final abundances decrease. 
While the higher $A$=94 production is a step in the right direction, a much higher abundance would be required for $A$=94 to be co-produced sufficiently to explain the solar abundance of $^{94}$Mo (Fig.~\ref{fig:abundance}). The change in $A$=94 production is due to the reduction of the $S_\mathrm{p}$ of $^{95}$Ag (from 1040 keV to 790 keV), which is a result of the new mass extrapolation for $^{95}$Ag. 
 
\begin{figure}
\includegraphics[angle=0,width=80mm]{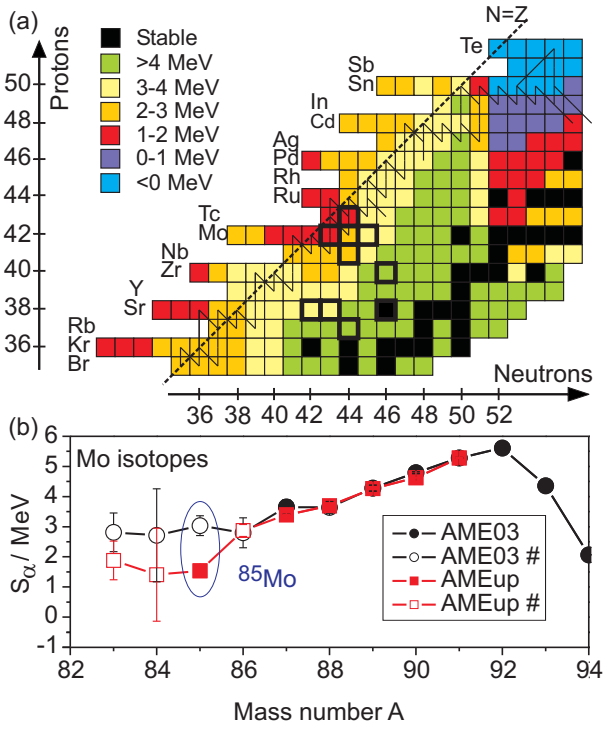}
\caption{\label{fig:alphaisland}(Color online) (a) Alpha separation energies, $S_\alpha$, for AMEup and the $rp$ process path. Nuclides addressed in this work are indicated by a thicker frame. (b) $S_\alpha$ of molybdenum isotopes for the AME03 (circles) and AMEup (squares). Filled symbols indicate measured values whereas open symbols indicate $S_\alpha$ where at least one of the two masses is extrapolated (\#).}
\end{figure}

The results also demonstrate for the first time the existence of a pronounced island of very low $\alpha$-separation energies  ($S_\alpha$)  for neutron-deficient Mo isotopes, starting with $^{85}$Mo for which our measurement now
provides an experimental $S_\alpha$ value  (Fig.~\ref{fig:alphaisland}).  A similar situation is found for the Tc isotopes
with our new  $S_\alpha$ value for $^{87}$Tc. This opens up the possibility for the 
formation of a ZrNb cycle induced by large $^{84}$Mo($\gamma$,$\alpha$) or $^{83}$Nb(p,$\alpha$) reaction rates. 
The existence of such a cycle had been discussed before  \cite{Schatz_1998} based on the low $S_\alpha$ around $^{84}$Mo predicted by the Finite Range Droplet mass Model (FRDM) \cite{Moeller_1995}. However, the FRDM is known to underpredict $S_\alpha$ severely in other isotopic chains, and the extrapolations provided by AME03 did not show such an effect. It was therefore concluded that the existence of a ZrNb cycle is an artifact of the FRDM. Because of the new measurements this view has to be revised.

\begin{figure}
\includegraphics[angle=0,width=80mm]{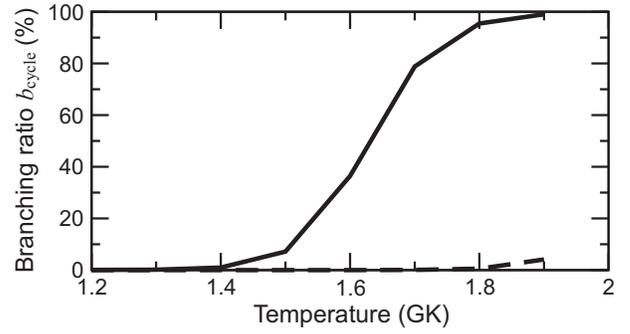}
\caption{\label{fig:cyclebranch} Fraction of reaction flow branching into a ZrNb cycle for AMEupx (solid line) and AME03x (dashed line) as a function of temperature. AME03 and AMEup show no significant cycling.}
\end{figure}

The existence of a ZrNb cycle in the $rp$ process is an important question. For a given $S_\alpha$ value of $^{84}$Mo
the cycle will form at a sufficiently high temperature, effectively providing an upper temperature limit for any $rp$ process along the proton drip line to produce nuclei beyond A$=$84, including the light p-nuclei in the A$=$92-98 
mass region. In order to explore whether our new results move this temperature limit into an interesting range, calculations in the Zr-Mo region were performed using a small test network with an initial $^{82}$Zr 
abundance.
The degree of cycling was determined as a function of temperature and density. 
As a measure of the degree of cycling, the fraction of nuclei $b_{\rm cycle}$ that end up in the $N$=40 isotonic chain via $^{84}$Mo($\gamma$,$\alpha$) or $^{83}$Nb(p,$\alpha$) and are not escaping the cycle via  $\beta^+$ decay into $N$=43 isotones were used. 
In order to explore the sensitivity of the results to nuclear masses additional mass data sets AME03x and AMEupx were created. Compared to AME03 and AMEup, a few masses of AME03x and AMEupx are changed by one standard deviation to the most favorable conditions for a cycle (low $S_\alpha$ and low $S_\mathrm{p}$ for $^{84}$Mo) to form.
For each set of masses all forward and reverse reaction rates were recalculated using the statistical model code NON-SMOKER$^\mathrm{WEB}$ version v5.8.1w \cite{Rauscher_2010}. 
Compared to AME03, calculations with AMEup, AME03x, and AMEupx increase the $^{83}$Nb(p,$\alpha$) reaction rate by factors of 40, $3\times 10^{3}$, and $8\times 10^{4}$ at 1.7 GK, respectively, while leaving the $^{83}$Nb(p,$\gamma$) reaction rate largely unchanged. Nevertheless, even for AMEupx, the  $^{83}$Nb(p,$\gamma$) reaction rate is still three orders of magnitude larger than $^{83}$Nb(p,$\alpha$). Regardless of the masses, a cycle can therefore only form when  $^{83}$Nb(p,$\gamma$) is suppressed by a large inverse $^{84}$Mo($\gamma$,p) reaction, or when $^{84}$Mo($\gamma$,$\alpha$) becomes significant. Only for AMEupx one finds  a significant $b_{\rm cycle}$ for realistic temperatures up to 2~GK (Fig.~\ref{fig:cyclebranch}). For higher temperatures, the radiation pressure in X-ray bursts would lead to  expansion and cooling, and heavy nuclei such as $^{84}$Mo would tend to be destroyed by photodisintegration. Both $^{84}$Mo($\gamma$,$\alpha$)  and $^{83}$Nb(p,$\gamma$) play a comparable role for the degree of cycling and the result does not depend significantly on density. While for high density more material is pushed past $^{83}$Nb, reducing the role of $^{83}$Nb(p,$\gamma$),  $^{84}$Mo($\gamma$,$\alpha$) turns out to be sufficiently fast to compensate, keeping the total $b_{\rm cycle}$ largely unchanged. A density range from 10$^5$ to $5 \times 10^{6}$~g/cm$^3$ was explored.

Despite the significant cycling found for AMEupx masses at high temperatures, calculations with the full X-ray burst model, which reaches peak temperatures of 2~GK, show that a cycle does not occur. The reason is that at the required high temperatures the reaction sequence already stops at $^{56}$Ni. Temperatures that allow processing beyond $^{56}$Ni are only reached during burst cooling and are lower than required to form a ZrNb cycle. The formation is nevertheless a possibility in an environment where the temperature is rising slowly enough to enable the $rp$ process to proceed past $^{56}$Ni before reaching high temperatures. Another possibility would be an $rp$ process with seed nuclei beyond $^{56}$Ni.

In summary, the masses of ten proton-rich nuclides were measured, among them  for the first time the nuclides  $^{85}$Mo and $^{87}$Tc. These are the heaviest $N$=$Z$+1 nuclides measured so far, extending the previous measurements at other facilities. Thus this work marks a milestone towards high-precision mass measurements of the heavy $N$=$Z$ nuclides, e.g. $^{80}$Zr and $^{84}$Mo. The results show a shift of the mass surface towards less bound nuclei. The $S_\mathrm{p}$ of $^{87}$Tc was experimentally determined for the first time and leads to large changes in the final abundance of $A$=86 produced in the $rp$ process. The abundance of $A$=94, which is the progenitor for production of $^{94}$Mo, was found to be sensitive to the $S_\mathrm{p}$ of $^{95}$Ag. First evidence is found of the existence of an island of low $S_\alpha$ in the Mo region which is traversed by the $rp$ process. 
This results in a revised, much lower prediction of the  $S_\alpha$ value of $^{84}$Mo. Within uncertainties this opens up 
the possibility for the existence of a ZrNb cycle, which would impose an upper temperature limit for the $rp$ process
to synthesize nuclei beyond A$=$84. In order to quantify this temperature limit and to explore to which degree such a cycle poses a limitation to current models of the $rp$ process precise mass measurements of  $^{80,81}$Zr, $^{83}$Nb and $^{84}$Mo are required.
Both the low $S_\mathrm{p}$ of $^{87}$Tc and the low $S_\alpha$ in the Mo region show that nuclear physics 
uncertainties in $rp$ process calculations can be surprisingly large and that experiments are needed to 
put astrophysical models on a solid foundation. 


\begin{acknowledgments}
This work was supported by the Helmholtz Association and GSI (VH-NG-033), BMBF (06GI185I, 06ML236I, 06GF9103I), Max-Planck Society and Swiss NSF 200020-122287. HS is supported by NSF grants PHY0822648 (JINA) and PHY0606007.

\end{acknowledgments}


\end{document}